# The Efficiency Challenges of Resource Discovery in Grid Environments

MAHDI MOLLAMOTALEBI[1], RAHELEH MAGHAMI[1],
ABDUL SAMAD ISMAIL[2], and ALIREZA POSHTKOHI[3]

[1]*Department of Computer Science, College of Engineering, Buinzahra Branch, Islamic Azad University, Buinzahra, Iran*
[2]*Faculty of Computing, Universiti Teknologi Malaysia, Skudai, Malaysia*
[3]*Department of Electrical Engineering, Shahed University, Tehran, Iran*

*Resource discovery is one of the most important services that significantly affects the efficiency of grid computing systems. The inherent dynamic and large-scale characteristics of grid environments make their resource discovery a challenging task. In recent years, different approaches have been proposed for resource discovery, attempting to tackle the challenges of grid environments and improve the efficiency. Being aware of these challenges and approaches is worthwhile in order to choose an appropriate approach according to the application in different organizations. This study reviews the most important factors that should be considered and challenges to be tackled in order to develop an efficient grid resource discovery system.*

## INTRODUCTION

The high-data and process-intensive applications (such as climate forecasting, protein folding, financial analysis, and so on) usually cannot be handled by a single machine. Grid computing allows carrying out such applications through collecting and collaborating hardware and software resources. The grid system is an infrastructure that provides dependable, consistent, and pervasive access to high-end computational capabilities (Foster and Kesselman 1999). The aim of grid computing is stated as sharing some resources to solve a problem in a dynamic environment formed by multiple virtual organizations (Foster, Kesselman, and Tuecke 2001).

Resource discovery is an important service that significantly affects the efficiency of grid systems. It is the mechanism of collecting and indexing the grid resources information and locating them as required. Grid nodes share their resources, and users issue the resource queries to access the required resources. The resource discovery system finds the required resources and reports their location to the requesting user. The performance of a grid system depends mainly on the efficient use of resources, but the characteristics of a grid environment make the resource discovery a time- and message-consuming process, which may decrease the performance of the entire grid system (Cokuslu et al. 2010). Resource requests in grid should be satisfied quickly and completely, otherwise the requesting application may be failed. Therefore an efficient resource discovery system is indispensable.

Previous resource sharing systems, such as peer-to-peer (P2P), used different techniques for resource discovery, but resources in a grid environment are heterogeneous types of hardware/software (e.g., CPU cores, storage systems, network bandwidths, files, services, etc.). Moreover, grid networks have some particular features such as dynamicity, ubiquity, and scalability. These specific features of grids and their resources make the grid resource discovery a challenging task (Vanthournout et al. 2005). Grid resources can be classified by the rate of their changes: (1) *Fixed resources* have a static position and characteristics such as a printer with a fixed resolution. (2) *Replicable resources* are fixed resources with replicated files. The resource discovery mechanism can report the address of one of the copied files. (3) *Mobile resources* can be fixed or replicable. They are variable in position, such as a moving laptop. (4) *Dynamic resources* are in a fixed position with varying values or characteristics such as current available memory. A resource can also be a combination of the above classifications, such as a mobile and dynamic resource (Vanthournout et al. 2005).

There are two main search approaches in grid resource discovery:

1. *Blind search*, which is based on the flooding mechanism. In this approach, grid nodes do not have information about existing resources, nodes, and their location. Resource requests are propagated to a sufficient number of nodes in order to satisfy the request. This type of search typically increases the network traffic, and subsequently, the performance of the search will be decreased significantly. Gnutella and random-walks are two popular methods that use the blind search (Tsoumakos and Roussopoulos 2003a,b). Some limiting factors, such as time-to-live (TTL), are used to avoid time wasting and to control the number of messages in such methods.

2. *Informed search* acts as directed and uses the historical information from the previous successful search results. Such information is indexed in centralized or distributed directory services to assist the search for a requested resource. These methods attempt to choose appropriate neighbors, based on the probability of success, to forward the resource query. Considering the directed search in such methods, their message loads are typically limited and they are preferred to be used in large-scale grid environments. Many search methods, such as CAN (Tang et al. 2003), Chord (Stoica et al. 2003), and others act as informed to speed up their search process while keeping the message loads low. The shortcoming of most informed search methods is the high maintenance costs of the resource information indices, especially in high-dynamic conditions.

In this article, we review the most important challenges and considerations in grid resource discovery, and how different methods confront these challenges. The rest of this article is organized as follows: "Resource Discovery Approaches" presents the main categories of resource discovery approaches in grid environments. In "Factors and Challenges of Grid Resource Discovery," we review challenges and factors associated with the grid resource discovery and how different methods tackle these challenges; "Conclusions" concludes our work.

## RESOURCE DISCOVERY APPROACHES

To investigate the behavior of different resource discovery approaches against the existing challenges of grid environments, the main resource discovery approaches are introduced in this section. Several resource discovery methods have been proposed for various types of distributed systems, but the particular characteristics of grid environment make the resource discovery a challenging issue. The resource discovery approaches in grid environments fall into one of the following categories.

In a centralized approach (see Figure 1), one or a limited number of central servers index(es) the resource information of an entire grid.

Each grid node includes some resources to share. The node publishes its resource information toward the central server. Also, once the resource information of the grid node is changed, for instance when its resource is allocated to a requesting application, the new information of the resource will be sent by the node directly to the central server. In some cases, these updates might be sent toward the central server periodically to control the message overhead of the network, but in dynamic environments, such periodic updates make the resource information of index servers unreliable. When a node needs some resources to handle its applications, it sends the resource query to the central server. The central server analyzes the received query, finds the appropriate resource owner in its database(s), and then returns the results to the requester node. Centralized methods are able to find the required resources quickly and accurately because they benefit from integrated databases.

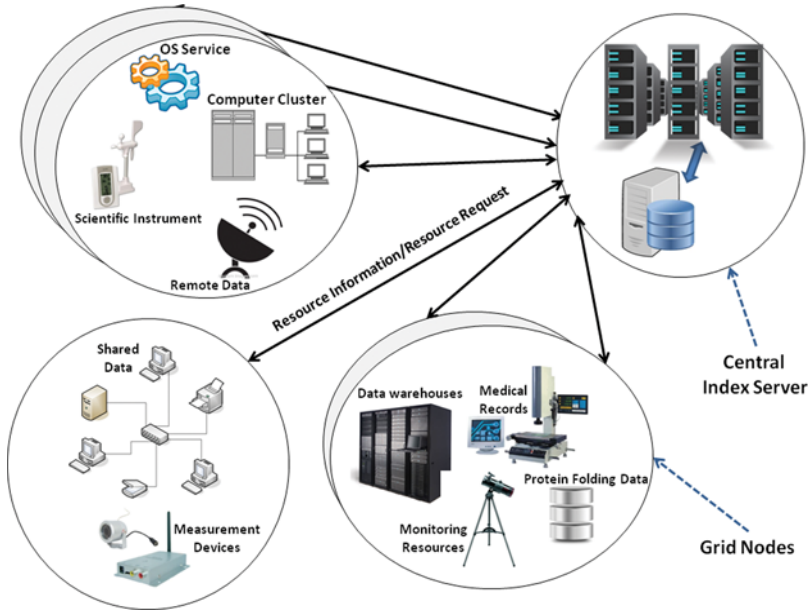

**FIGURE 1** Centralized resource discovery.

Most centralized methods, such as found in Yu et al. (2003) and Kaur and Sengupta (2007), exploit the existing information directory protocols, lightweight directory access protocol (ldap) and universal description discovery and integration (UDDI), to handle their resource information index (Cokuslu et al. 2010).

In the hierarchical/tree approach, which is shown in Figure 2, the resource information is partially distributed to multiple locations as hierarchical.

In the hierarchical tree, grid nodes publish their resource information to their parent located in upper level (Cokuslu et al. 2010; MollaMotalebi et al.

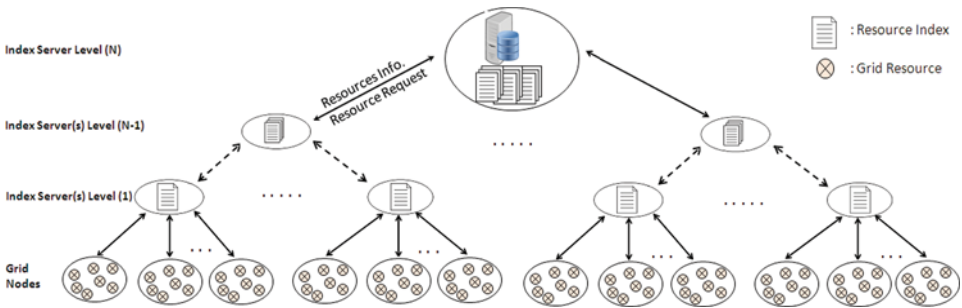

**FIGURE 2** Hierarchical resource discovery.

2011). Also, each grid node handles the queries related to its descendants. In some cases, grid nodes do not maintain the metadata of all their descendants because of security considerations or memory-space limitations. The hierarchical approach is prevalent in many current resource management systems such as Globus Toolkit, gLite, etc. (Mastroianni et al. 2008).

As shown in Figure 2, index servers at level 1 of the hierarchy are directly connected to the resource owner nodes. They gather the resource information and forward it to their parent to be indexed. Thus, the upper-level indexing nodes are busier and contain the information about more resources in the network; they have a more important role in the resource discovery system and necessitate more maintenance costs, especially in dynamic conditions.

In a P2P-based approach, as shown in Figure 3, connections between the grid nodes can be viewed as a graph. The P2P structure dictates a distributed and cooperative network design without central supervision and the information of resources is not kept on a central server. Each node has the role of both client and server and acts as autonomous and self-organized. Grid applications send their resource query message to their neighbors. The neighbors subsequently forward the received queries to appropriate neighbors in order to reach the required resource location.

P2P-based approaches can be categorized as unstructured and structured. In the unstructured P2P-based approach, grid nodes are connected to each other without predefined constraints. The resource discovery is typically handled by using flooding-based mechanisms in such approaches.

However, structured P2P-based approaches are organized by some specific constraints in terms of the indexing and network topologies. They may utilize distributed hash tables (DHT) to index the resource information (Trunfio et al. 2006, 2007). Furthermore, grid nodes may record the information of previous successful search results to be exploited in the next search processes (Ranjan et al. 2008).

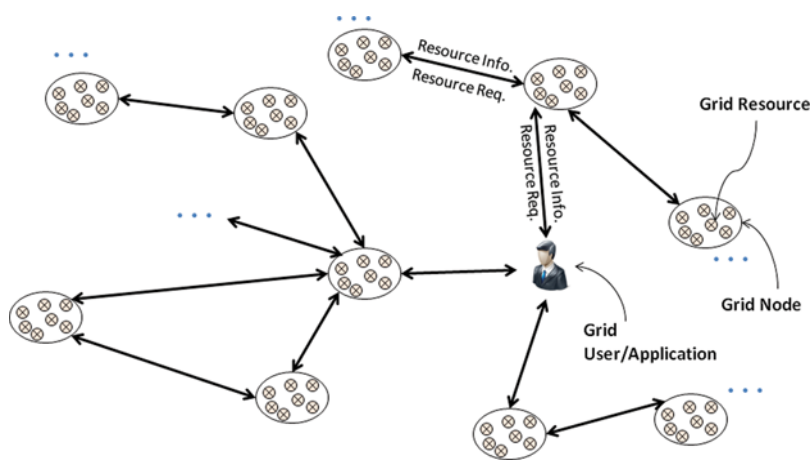

**FIGURE 3** P2P-based resource discovery system.

The super-peer-based approach, which is shown in Figure 4, attempts to balance between the inherent accuracy of centralized systems and the autonomy and load-balancing offered by P2P-based systems (Mastroianni et al. 2007).

In this approach, two types of nodes exist as super-peer and regular nodes. Each super-peer node gathers and indexes the resource information of a limited number of regular nodes. The regular nodes send the resource queries to their related super-peer node. The super-peer nodes are related to each other as P2P, and they are responsible to find the resources required by their regular nodes (Mastroianni et al. 2005; Trunfio et al. 2007).

The regular node sends the resource query to its local super-peer. If the local super-peer finds the desired resources within its local index, it forwards the requesting node to the owner of the desired resource; otherwise, the super-peer node forwards the query message to its neighboring super-peer nodes to traverse the grid network. This approach is almost scalable because a large-scale grid is handled as several interconnected small-scaled grids. Therefore, the super-peer approach benefits from the advantages of centralized and P2P structures while decreasing the probability of single-point-of-failure and bottleneck problems.

Grid resource discovery approaches can also act based on the agents, see Figure 5. An agent (especially a mobile agent) is a piece of software with particular characteristics such as autonomy, ongoing execution, mobility, and reactivity. It can be exploited in grid environments to increase the scalability and flexibility of resource discovery (Hameurlain et al. 2010).

Each agent can be mapped into one grid node to keep its resource information. It also is able to navigate the network to find the requested resources. The autonomy of agents enables them to act as intelligent. They are able to change their behavior according to different conditions during the search process (El Mahmoud et al. 2008; Yan et al. 2007).

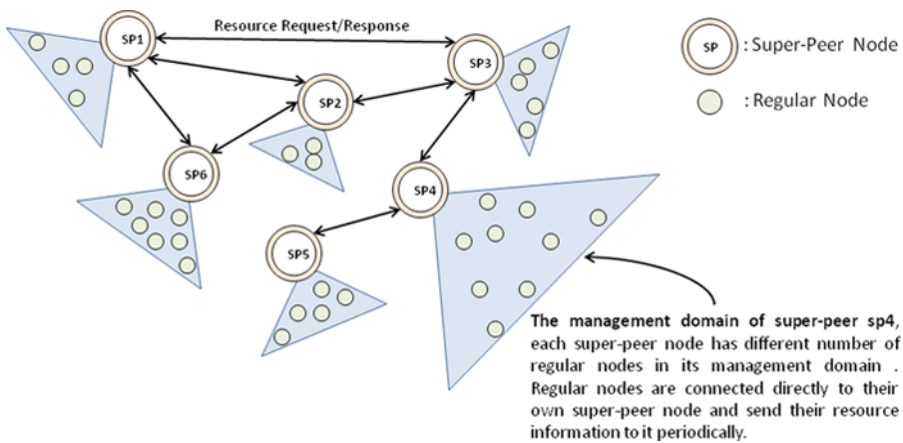

**FIGURE 4** Super-peer-based resource discovery system.

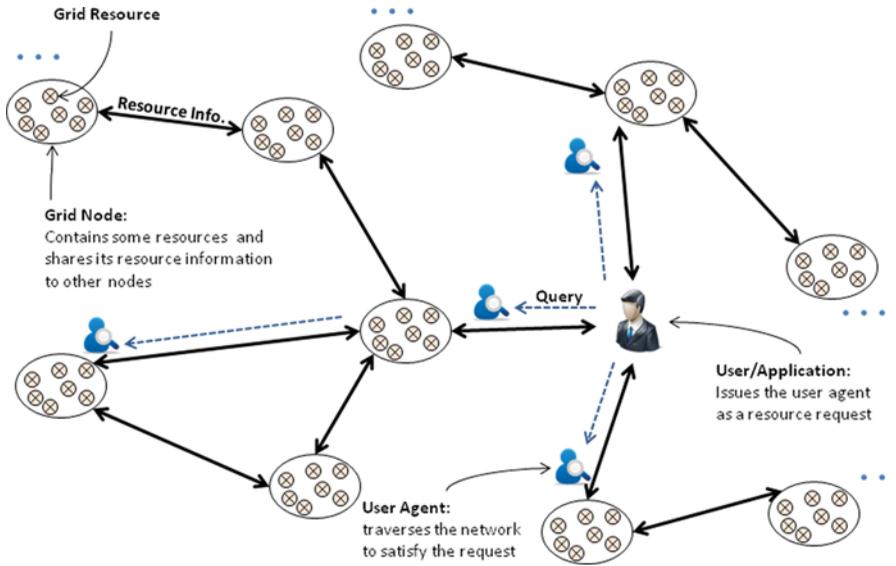

**FIGURE 5** Agent-based resource discovery system.

## FACTORS AND CHALLENGES OF GRID RESOURCE DISCOVERY

Several resource discovery models and algorithms have been developed in P2P and grid environments. With regard to the characteristics of grid environments, the resource discovery deals with some challenges/considerations, which are presented in this section. The stratagems utilized by the resource discovery approaches to tackle these challenges are also presented in this section.

### Dynamicity and Heterogeneity

Grid environments are usually highly dynamic in terms of the users, resource values, and resource availability. Also, grid resources are heterogeneous with a broad spectrum of resource types such as computing devices, storage, supercomputers, and clusters (Cokuslu et al. 2010). Furthermore, they are provided by organizations with different administrative domains and access policies, which influence grid application performance (Mastroianni et al. 2003). Heterogeneity also arises from various resources available within different scopes. For instance, one software package might be able to be handled by only some particular operating systems, whereas its data can be extracted from different data management systems (e.g., relational databases, plain files, etc.; Mastroianni et al. 2004). The heterogeneous environments deal with challenges to control and balance the resource's access and deployment across the network (Randles et al. 2010).

Grids are dynamic from different points of view: (1) The number of users and nodes participating in the grid fluctuates over time (Abdelkader and Broeckhove 2008). (2) The resource requirements of grid applications change unexpectedly (Zhengli 2010). (3) Some resource values (such as CPU load and available memory) vary rapidly over time (Zhengli 2010). (4) The network conditions (e.g., access policies and bandwidth) might vary repeatedly (Slimani and Yagoubi 2007).

The joint process typically necessitates some registration and authentication constraints because grids are formed by different virtual organizations with their own rules and limitations. But resources may leave the grid at any time with a defined leave process or even without any notification; it can take place because of a node failure. Node failure might be caused by a network communication problem or failure in the node itself. Thus, it can lead the resource discovery process to incorrect results. As an example, the search process might return some resources that are inaccessible in the grid network. Resource information replication can speed up the search process and make it more accurate because queries can be answered by resource repositories close to them; however, in dynamic conditions, the cost of replication is excessive (Ranjan et al. 2008; Trunfio et al. 2007). Furthermore, some resource discovery methods exploit caches in order to benefit from historical information related to performed search, but dynamicity of grid environment also negatively affects cache usefulness because cached information quickly becomes stale (Iyengar et al. 2004).

Most centralized and hierarchical resource discovery methods (Ramos and de Melo 2006; Kaur 2007; Molt et al. 2008) are not adapted to dynamic grid environments. Unstructured P2P-based methods (Iamnitchi and Foster 2004; Filali et al. 2008) are adapted to dynamic conditions because their indexing nodes are distributed on the network and resource changes can be updated quickly. Structured P2P-based methods (Cai et al. 2003; Oppenheimer et al. 2004; Tali et al. 2007) poorly support the dynamicity due to high costs of keeping the distributed hash tables up to date. Also, super-peer-based methods (Mastroianni et al. 2005) act weakly in dynamic grid environments because failure of a super-peer node causes some regular nodes' information to be lost. Some super-peer-based methods (Puppin et al. 2005) attempt to reduce this problem by using redundant super-peer nodes, however, the redundancy imposes overloads on the system. Agent-based methods are roughly adapted with dynamic conditions because the agents are flooded as parallel to the network to gather resource information simultaneously, but the indexed information may not be up to date because of periodic index updates.

Therefore, dynamic grid environments increase the message and processing loads of grid nodes. Among different approaches for grid resource discovery, the unstructured P2P-based and agent-based are more adapted to dynamic conditions and are preferred to be employed in such environments.

Scalability

The scale of grid networks, in terms of number of resources, geographical spread, and number of users/applications can be small or large, but grid environments usually tend to be large in scale (Comito et al. 2005, Hameurlain et al. 2010). Therefore, resource discovery methods for grid environments should be able to match with such potential large scales (Foster et al. 2001).

The methods that utilize a flooding mechanism to disseminate and search the resource information (e.g., Gnutella), are not scalable due to massive message loads. Centralized grid resource discovery methods typically are not scalable because of a potential bottleneck problem in the central index node in large-scale conditions. In large scales, the number of messages that should be handled by central servers dramatically increases and it can cause the whole grid system to act inefficiently or even fail. In hierarchical methods, upper-level indexing nodes need more memory space to maintain information of their descendant resources in a large-scale grid. Also, management to quickly search the required resource among the huge amounts of information is a challenging task, especially in upper levels of hierarchical structure.

P2P-based resource discovery methods remove the bottleneck in large-scale environments; however, unstructured P2P-based methods are not scalable due to high message loads in large scales. False positive error is another challenge in most of the unstructured P2P-based system that is caused by using limiting factors such as TTL for query message forwards (Foster et al. 2001; Tsoumakos and Roussopoulos 2003a,b). Most of the structured P2P-based methods (Cheema et al. 2005; March et al. 2007; Albrecht et al. 2008) use distributed hash tables and act properly in large-scale environments because they typically need message loads lower than the order of grid nodes' growth (Hameurlain et al. 2010). Furthermore, some techniques (Sahota et al. 2009; Shen 2008; MollaMotalebi et al. 2011) group the nodes and/or resources into clusters in order to improve the scalability.

The super-peer-based methods are more scalable than the unstructured P2P-based because they divide a large-scale grid to multiple small scales (Mastroianni et al. 2008). Super-peer-based methods still deal with the bottleneck problem at the point of indexing nodes in large scales. In the agent-based methods, the flooding is used to disseminate the resource information or resource request messages among the network nodes. With regard to uncontrolled high numbers of flooding agent messages in large scales, such methods deal with traffic overloads and delay during the search process. Taking into account that grid environments typically tend to large scales, it is critical to consider scalability preparations in designing the resource discovery systems. Among current resource discovery approaches, the structured P2P-based and super-peer-based are roughly adapted to large-scale grid environments.

Reliability/Stability

With regard to the dynamic nature of the grid environment, the nodes may join, leave, or fail at any time. Such events for the indexing nodes can cause some grid resources to be inaccessible. Grid resource discovery methods attempt to be tolerant against such problems by applying the reliability considerations (Puppin et al. 2005). The main factors dealing with reliability/stability of grid resource discovery approaches are bottleneck, single point of failure, false positive error, unbalanced loads, and lack of up-to-date indexed information.

The bottleneck occurs when a large number of resource queries are passed to the limited number of indexing nodes. In this case, the indexing nodes deal with the message and processing overloads. Also, single point of failure refers to the nodes that contain some important information so that other parts of the network depend on them. Thus, absence of such important nodes makes the dependent nodes unable to share and discover the resources. False positive error happens when a resource discovery system is not able to traverse the entire grid network because of some limitations (such as TTL) and it returns wrong responses for the search queries. Unbalanced loads occur when some indexing nodes in the network are so busy and some others are idle based on the memory usage, processing operations, bandwidth traffic, message forwarding, and so on. This causes negative performance of the entire grid. Lack of up-to-date indexed information occurs from late or periodical gathering of the resource changes. It makes the indexed information to be untrustworthy because the resources may be allocated to an application or released by it during the period of update delay, but the resource information changes are not updated in the indexing node.

Some grid developers prepare backup or redundant nodes (Hameurlain et al. 2010) for important indexing nodes to keep the system more reliable. In this case, update or replication operations impose some overheads in terms of the bandwidth, storage, and processing loads to the system (Cokuslu et al. 2010). Single point of failure and bottleneck is a common problem in centralized methods (Trunfio et al. 2007) because a large part of the network depends on central servers and their functions (Molt et al. 2008). The hierarchical approach encounters the single point of failure especially in the upper levels of hierarchy. Moreover, delay of updates in upper-level indexing nodes impacts negatively the reliability of hierarchical structure (Mastroianni et al. 2008).

Taking into account the heterogeneous and dynamic characteristics of grid environments, grid resource indexes typically become unbalanced.

Besides the centralized approach, the hierarchical approach also deals with the unbalance challenge because index servers belonging to different levels need to handle different computation and message loads (Mastroianni et al. 2008). Thus, some nodes might cause the overhead of data and processing loads, while some others' capabilities are wasting. This affects negatively the performance of the whole system. Furthermore, unbalanced load implies unbalanced importance of roles so that some nodes may have a critical and some others a negligible role in the network. It necessitates a resource discovery system to keep the indexing and processing loads as balanced as possible.

Structured P2P-based methods (Albrecht et al. 2008, Trunfio et al. 2007) are not fault tolerant because leaving their index nodes necessitates reorganizing the resources and index tables. In addition, they are not load balanced because the indexing nodes containing the information of popular resources deal with high overloads; however, some efforts have been made to improve their load balancing (Chen et al. 2005; Cai and Hwang 2007; Ranjan et al. 2007). On the other hand, they do not deal with single point of failure and false positive errors. Unstructured P2P-based methods provide more autonomy for the participants of a grid and they act as load balanced, however, they still are not reliable because of the false positive errors caused by using message flooding and TTL.

Super-peer-based methods are weakly reliable because of single point of failure in the point of super-peer nodes, especially in dynamic conditions. Also, in large-scale environments, false positive error is highly probable. Furthermore, because the super-peer nodes are updated periodically, the lack of fresh resource information is foreseeable. Agent-based systems use the flooding mechanism for agent messages caused by the false positive error. Also, if they do not use TTL, they fall into the message overload and high latency to reach the result. Therefore, among current resource discovery approaches, the structured P2P-based and super-peer-based approaches are relatively reliable.

## Diversity of Resource Requests

*The* resource discovery system is responsible for locating the required resources of grid applications. The resource queries issued by grid applications may be in different forms. The queries should denote the resource requirements as accurately as possible. For instance, requesting a processing resource (e.g., CPU) without additional specifications is not adequate because some resources may be allocated to requesters that are higher than their requirements and it affects negatively the efficiency of the grid system. Resource queries may be formed as the following:

1. Multiattribute queries: A multiattribute query includes multiple requested resources. For instance, a query might request two computing and one storage resources. The multiattribute query involves a set of subqueries

　　on individual resources, and in some cases each subquery can be a multi-attribute query too.
2. Ranged-attribute queries: sometimes a query requests a ranged numeric, or string values, of a resource (e.g., 1GHZ <CPU Intel <4GHZ) (Talia et al. 2007).
3. Dynamic-attribute queries: attributes of resources may be static or dynamic. Static attributes do not change frequently, such as "CPU clock rate." Dynamic-attribute queries request the resources that their values vary on time, such as "CPU usage" or "available memory" (Ranjan et al. 2008; Hameurlain et al. 2010).

　　Today's resource discovery methods attempt to support all of the aforementioned query types. Some methods utilize space-filling curves, znet (Shu et al. 2005) and CISS (Lee et al. 2007), to convert a multiattribute query to different single-attribute queries. Squid (Schmidt and Parashar 2004) uses statistical information to make the space partitioned. Chawathe et al. (2005) forms a prefix hash tree on any Distributed Hash Table (DHT). Network-R-tree (NR-tree) (Liu et al. 2005), which is in the form of super-peers, offers approximate answers for ranged-attribute queries and maintains multiple indices. Virtual Binary Index Tree (VBI-tree) (Ooi et al. 2006) has a P2P structure and employs a balanced tree to perform the centralized mapping, such as Rectangle Tree (R-Tree), eXtended node Tree (X-Tree), Similarity Search Tree (SS-Tree), and Metric Tree (M-Tree).

　　Supporting dynamic attribute queries in grid resource discovery approaches is more challenging than multiattribute and ranged-attribute queries because it necessitates having fresh and up-to-date information about the resources at any time. The dynamic nature of grid environments intensifies handling such queries because the update process is typically accomplished periodically in order to keep the message and processing loads low.

　　Centralized and hierarchical approaches support range-attribute and multiattribute queries because they usually utilize limited and integrated databases to index and search the resource information, which allows them to do flexible, quick, and accurate searches. But they cannot support dynamic-attribute queries efficiently because of the delay of indexing nodes' updates especially in large-scale and dynamic environments. P2P-based and agent-based methods typically support all of the query types, however, the structured P2P-based methods may be too complex to update DHTs, and some of them are not able to support dynamic-attribute queries. Super-peer-based methods (Marzolla et al. 2005; Mastroianni et al. 2005; Puppin et al. 2005) support multiattribute and ranged-attribute queries but they cannot support dynamic-attribute queries because of periodic collecting of the resource information. Moreover, agent-based methods support all types of queries. Therefore, almost all of resource discovery approaches support the different types of resource queries; however, for the dynamic-attribute queries, they are faced with some difficulties.

Quality of the Results

Grid applications expect to achieve the highest quality of results. Quality of results in a grid resource discovery system can be measured by different aspects. The most important factors to measure the quality of results are as follows (Yang and Garcia-Molina 2002; Garcia-Molina and Yang 2003; Tsoumakos and Roussopoulos 2003; Trunfio et al. 2006; Reinhard and Tomasik 2008):

1. *Number of the results* is one of the factors that determines the quality of the search results. Having a larger number of results implies that the user has a chance to choose the most appropriate result among the set of found resources according to its priorities (e.g., choosing the geographical nearest resource, choosing based on the network bandwidth and so on).
2. *Success rate* implies the average rate of getting the results in the first attempts of query issues. Apparently, a system that is able to find the required resources with fewer search retries is more appropriate. Typically, informed search methods that use directed search (e.g., methods based on previous search records and methods based on resource categorization) have more chances to locate the required resources in first attempts.
3. *The accuracy of the results* measures how many of discovered resources are relevant and how relevant they are to the issued resource request. Resource discovery systems should be able to locate the accurate matched resources to the application's requirements. As mentioned in a previous section, the appropriate resource query also can affect the accuracy of results.
4. *Response time* implies the time between issuing the resource query and returning the search result to the initial requester. Considering the large scale (in terms of the number of nodes, number of resources, and geographical distances) nature of grid environments, quickly locating the required resources is a critical challenge for resource discovery approaches.

Centralized and hierarchical methods are able to find required resources with the best quality only on small scales. This is because they typically use centralized and integrated databases to index and search the resource information that provides the accurate and fast search tasks (Filali et al. 2008; Mastroianni et al. 2008; Cokuslu et al. 2010). Unstructured P2P-based methods usually find a large number of results, but with less accuracy. Structured P2P-based and super-peer-based methods provide high quality results; although, in terms of the number of results and success rate, they are not as effective as other approaches, in large-scale environments, they are preferred and provide the highest quality of results (Trunfio et al. 2007; Mastroianni et al. 2008). Considering the main goal of the resource discovery, which is to achieve the set of required resources location, the quality of results is considerable. Overall, and with regard to the inherent large scale of grid environments, the structured P2P-based and super-peer-based approaches are more appropriate in terms of the quality of results.

Maintenance Cost

Maintenance cost is a considerable factor for the grid resource discovery's efficiency. It can be measured by the overheads imposed to a grid system in terms of some aspects such as processing loads (e.g., routing), storage usage (e.g., database updates), bandwidth usage (e.g., message loads), number of nodes dealing with resource information dissemination, or discovery, usage frequency of the nodes dealing with resource discovery, security considerations, and so on.

Though the main goal of grid resource discovery is to locate the required resources quickly and with high-quality results, they also should reach these goals with a reasonable maintenance cost. The resource discovery typically includes resource information dissemination, update, and search tasks. Each of these tasks necessitates some messages to be transferred in the network. For example, each node should notify the changed information of its resources to its neighbors. It is typically handled by broadcasting update messages. Considering the dynamicity and large scale of a grid environment, the number of update messages for different nodes in the network rises dramatically.

Though a flooding mechanism allows quick location of the desired resources, it necessitates very high message loads such that a large number of grid nodes deal with flooded messages. With regard to the bandwidth and processing limitations of grid nodes, flooding mechanisms can decrease the performance of an entire grid and increase the maintenance costs of grid nodes. Thus, the majority of methods tend to use informed (or directed) search mechanisms that limit the flooding domain of messages. Some solutions for doing an informed search include using the previous successful search results to make a decision for choosing the proper neighbors in order to forward query messages, record the resources shared by each node in hash tables, and so on. Although the informed search mechanisms reduce the number of flooded messages, they also need some message exchanges to keep the resource information indexes up to date.

Centralized methods impose high message and processing loads on the central index servers. Thus, the maintenance cost of the central servers is high. Hierarchical methods distribute the message and processing loads between multiple index nodes; however, upper-level indexing nodes still suffer from high maintenance costs caused by leave events and resource update messages of hierarchical descendant nodes. Furthermore, the security considerations should be applied on upper-level indexing nodes to protect them from bombarded messages sent by descendant nodes. One used strategy is that each node knows only its direct related parent, and the address of upper-level indexing nodes are hidden from grid nodes except for their children (Foster and Kesselman 1999; Yang and Garcia-Molina 2002).

**TABLE 1** Grid Resource Discovery Challenges/Factors in Different Approaches

| Factor approach | Dynamicity | Scalability | Reliability/ stability | Query types | Quality of results | Maintenance costs |
|---|---|---|---|---|---|---|
| **Centralized** | Weakly support due to bottleneck in dynamic conditions | Not supported due to potential bottleneck in central index nodes | Not reliable/ stable due to single point of failure and bottleneck in central servers | Support multiattribute and ranged-attribute queries | High quality in small-scale networks | High due to large amount of memory space and message complexity in central servers |
| **Solutions**: use in small-scale and low-dynamic environments, and prepare the backup index servers. Moreover, increase the central index servers to reduce the probability of single point of failure. ||||||||
| **Hierarchical** | Weakly support due to bottleneck problem in upper-level indexing nodes | Weakly support due to potential bottleneck in upper indexing nodes, and latency of resource information updates | Not reliable/stable due to potential single point of failure in index nodes, and unbalanced load | Weakly on dynamic attribute queries due to latency of resource information updates | High quality in small-scale networks | High due to large amount of memory space and security management in upper-level indexing nodes |
| **Solutions**: replication between index servers and utilizing redundancy for upper-level indexing nodes. Also, limiting the access of nodes to upper levels to keep high reliability and security. ||||||||
| **Super-peer-based** | Weakly support due to periodic update of resource information in super-peer nodes | supported because of dividing a large-scale network into multiple small-scales | Weakly supported due to potential single point of failure and false positive error | Support all types, weak in dynamic attribute queries in dynamic conditions | Provide higher quality of results relative to other approaches in large-scale networks | Tolerable because processing, message and memory costs are distributed between all the nodes |
| **Solutions**: controlling and adjusting the number of super-peers and their capacities depending on the network scale and dynamicity of resources, using more super-peers in large-scale networks and fewer super-peers in dynamic conditions. ||||||||

**TABLE 1** Continued

| Factor approach | Dynamicity | Scalability | Reliability/ stability | Query types | Quality of results | Maintenance costs |
|---|---|---|---|---|---|---|
| **P2P-based** | Structured P2P-based methods act poorly due to latency of applying the resource changes in distributed hash tables | Not scalable in the case of unstructured P2P-based due to huge growth of message loads in large scales and potential false positive errors | Supported because resource information is distributed, weak in the case of unstructured P2P due to false positive error | Support all types of queries | High quality, unstructured P2P-based deal with latency in large-scale environments | High, due to high costs of keeping the distributed hash tables as up to date |
| **Solutions**: optimizing the indexing strategies to keep the maintenance costs low. Furthermore, avoiding the use of unstructured P2P-based strategies in large-scale environments. | | | | | | |
| **Agent-based** | Supported due to using of multiple parallel mobile agents | Weakly scalable because of delay and huge number of messages | Not supported because of false positive errors | Support all types of queries | Low quality of results in large-scale environments | High, due to huge message and processing loads |
| **Solutions**: using in small-scale and low dynamic environments. Moreover, optimizing the indexing and locating strategies to keep the resource discovery system reliable and reaching higher quality of results. | | | | | | |

The structured P2P-based methods locate the resources quickly and with high quality, but most of them extremely suffer from high maintenance cost. It is due to management and update arrangements, which are necessary for indexing nodes such as DHTs. Unstructured P2P-based approaches do not use the hash tables, but as mentioned, they suffer from high message loads and number of nodes dealing with the resource discovery during the search process.

Because the super-peer-based methods divide a single large-scale network into multiple smaller scale ones, they could reduce the maintenance costs. It is also noteworthy that the maintenance cost in the super-peer approach depends on the number of regular nodes located within each super-peer. A higher number of regular nodes handled by super-peers speeds up the search process but with higher maintenance costs and vice versa. Therefore, there is a tradeoff between the quality of results and the maintenance cost in this approach.

The maintenance cost of agent-based methods is also high because of the exchange of huge numbers of agent messages during the search process that raises the processing loads of grid nodes. One solution for tackling this problem is to use cached information of previous found resources, but in dynamic environments it is not useful (Tsoumakos and Roussopoulos 2003; Trunfio et al. 2007).

Consequently, keeping the maintenance cost low in grid resource discovery systems is a challenging issue. This is because of some tradeoffs that exist between the goals of the resource discovery (e.g., fast and accurate location of the required resources) and maintenance cost metrics (e.g., low number of exchanged messages). Therefore, resource discovery approaches attempt to keep a balance between the maintenance costs and the quality of index and search processes.

Table 1 summarizes all the discussed challenges and factors related to the efficiency of resource discovery in grid environments.

# CONCLUSION

Resource discovery is one of the most important issues in grid computing. Considering the aim of the grid systems, which is to provide required resources for data intensive applications, efficient location of the required resources is a vital task. There are different factors (such as reliability/stability, response time, etc.) that should be considered when designing a grid resource discovery system. Moreover, with regard to inherent characteristics of grid environments (such as dynamicity, heterogeneity, etc.), providing an efficient grid resource discovery approach is a challenging task.

In this article, we reviewed some of the important factors that necessarily should be considered in a grid resource discovery system. We also mentioned the main challenges dealing with current grid resource discovery approaches and how the approaches tackle them.

By and large, centralized, hierarchical, unstructured P2P-based and agent-based approaches are suitable for low-dynamic and small-scale grid environments. In addition, the super-peer-based and structured P2P-based approaches can be used appropriately in dynamic and large-scale grid environments. Taking into account the tradeoffs among different factors and challenges, resource discovery systems can hardly tackle all the challenges concurrently. For example, there is a tradeoff between the accuracy of the result and dynamicity of the environment such that trying to find resources accurately in a dynamic environment necessitates increasing higher message loads and, subsequently, it affects negatively the maintenance cost.

Therefore, considering the nature of the grid environment and tradeoffs between different factors/challenges, it is almost impossible to design a resource discovery system that keeps all the aforementioned factors as efficient and able to tackle all the challenges concurrently. But taking into account the application of grid systems, it is possible to choose some factors to be considered in a resource discovery approach and to achieve balance among the discussed challenges.

*Resource Discovery in Grid Environments* 691Ranjan, R., A. Harwood, and R. Buyya. 2008. "Peer-to-Peer-Based Resource Discovery in Global Grids: A Tutorial." *IEEE Communications Surveys & Tutorials* 10, no. 2 (2008): 6–33.

Ranjan, R., L. Chan, A. Harwood, S. Karunasekera, and R. Buyya. "Decentralised Resource Discovery Service for Large Scale Federated Grids." In *Proceedings of the Third IEEE International Conference on e-Science and Grid Computing*. Washington, DC: IEEE Computer Society, 2007.

Reinhard, V. and J. Tomasik. "A Centralised Control Mechanism for Network Resource Allocation in Grid Applications." *International Journal of Web and Grid Services* 4, no. 4 (2008): 461–475.

Sahota, V., M. Li, M. Baker, and N. Antonopoulos. "A Grouped P2P Network for Scalable Grid Information Services." *Peer-to-Peer Networking and Applications* 2, no. 1 (2009): 3–12.

Schmidt, C. and M. Parashar. "Enabling Flexible Queries with Guarantees in P2P Systems." *IEEE Internet Computing* 8, no. 3 (2004): 19–26.

Shen, H. "SEMM: Scalable and Efficient Multi-Resource Management in Grids." Paper presented at the International Conference on Grid Computing and Applications (GCA), pp. 17–21. Las Vegas, Nevada, USA, July 14–17, 2008.

Shu, Y., B. C. Ooi, K.-L. Tan, and A. Zhou. "Supporting Multi-Dimensional Range Queries in Peer-to-Peer Systems." In *Proceedings of the Fifth IEEE International Conference on Peer-to-Peer Computing*. Washington, DC: IEEE Computer Society, 2005.

Stoica, I., R. Morris, D. Liben-Nowell, D. R. Karger, M. F. Kaashoek, F. Dabek, and H. Balakrishnan. "Chord: A Scalable Peer-to-Peer Lookup Protocol for Internet Applications." *IEEE/ACM Transactions on Networking* 11, no. 1 (2003): 17–32.

Talia, D., P. Trunfio, and J. Zeng. "Peer-to-Peer Models for Resource Discovery in Large-Scale Grids: A Scalable Architecture." In *Proceedings of the 7th International Conference on High Performance Computing For Computational Science*. Rio de Janeiro, Brazil: Springer-Verlag, 2007.

Tang, C., Z. Xu, and S. Dwarkadas. "Peer-to-Peer Information Retrieval Using Self-Organizing Semantic Overlay Networks." In *Proceedings of the 2003 Conference on Applications, Technologies, Architectures, and Protocols For Computer Communications*. Karlsruhe, Germany: ACM, 2003.

Trunfio, P., D. Talia, P. Fragopoulou, C. Papadakis, M. Mordacchini, M. Pennanen, K. Popov, V. Vlassov, and S. Haridi. "Peer-to-Peer Models for Resource Discovery on Grids." Technical Report TR-0028, CoreGRID, 2006.

Trunfio, P., D. Talia, H. Papadakis, P. Fragopoulou, M. Mordacchini, M. Pennanen, K. Popov, V. Vlassov, and S. Haridi. "Peer-to-Peer Resource Discovery in Grids: Models and Systems." *Future Generation Computer Systems* 23, no. 7 (2007): 864–878.

Tsoumakos, D. and N. Roussopoulos. "Adaptive Probabilistic Search for Peer-To-Peer Networks." In *Proceedings of the Third International Conference on Peer-to-Peer Computing*, (P2P 2003), 102–109. IEEE, 2003a.

Tsoumakos, D. and N. Roussopoulos. "A Comparison of Peer-to-Peer Search Methods." In *International Workshop on the Web and Databases*, 61. 2003b. Available at: http://www.ics.forth.gr/webdb03/index.htm.

692　　　　　　　　　　　　　*M. MollaMotalebi et al.*Vanthournout, K., G. Deconinck, and R. Belmans. "A Taxonomy for Resource Discovery." *Personal Ubiquitous Computing* 9, no. 2 (2005): 81–89.

Yagoubi, B. and Y. Slimani. "Task Load Balancing Strategy for Grid Computing." *Journal of Computer Science* 3, no. 3 (2007): 186–194.

Yan, M., Q. Yu-hui, W. Gang, and Z. Ju-Hua. "Study of Grid Resource Discovery Based on Mobile Agent." In *Proceedings of the Third International Conference on Semantics, Knowledge and Grid*. Washington, DC: IEEE Computer Society, 2007.

Yang, B. and H. Garcia-Molina. "Improving Search in Peer-to-Peer Networks." In *Proceedings of the 22nd International Conference on Distributed Computing Systems*. Washington, DC: IEEE Computer Society, 2002.

Yu, J., S. Venugopal, and R. Buyya. *Grid Market Directory: A Web Services Based Grid Service Publication Directory*. Grid Computing and Distributed Systems (GRIDS) Lab, Dept. of Computer Science and Software Engineering, The University of Melbourne, 2003.

Zhengli, Z. "Grid Resource Selection Based on Reinforcement Learning." In *2010 International Conference on Computer Application and System Modeling (ICCASM)*, V12–644–V12–647. IEEE Conference Publications, 2010.